Evaluation of Disorder Introduced by Electrolyte Gating through Transport Measurements in Graphene

Andrew Browning, Norio Kumada*, Yoshiaki Sekine, Hiroshi Irie, Koji Muraki, and Hideki Yamamoto

*NTT Basic Research Laboratories, NTT Corporation, 3-1 Morinosato-Wakamiya, Atsugi, Kanagawa, 243-0198, Japan*

Email: kumada.norio@lab.ntt.co.jp

We evaluate the degree of disorder in electrolyte gating devices through the transport measurements in graphene. By comparing the mobility in ion- and standard metal-gated devices, we show that the deposition of the ionic liquid introduces charged impurities, by which the mobility in graphene is limited to $3 \times 10^3$ cm$^2$/(Vs). At higher temperature ($> 50$ K), phonons in the ionic liquid further reduce the mobility, making its upper limit $2 \times 10^3$ cm$^2$/(Vs) at room temperature. Since the degree of disorder is independent of the base material, these results are valuable towards understanding disorder effects in general devices using electrolyte gating.

Electrolyte gating has been a prominent topic lately for its widespread applications. One of its distinctive features is the large capacitance (~10 µF/cm$^2$) due to the formation of the electric double layer (EDL). This large capacitance, together with the flexibility, transparency, and facile processing of ionic liquids, is useful for soft electronic devices.[1-4] Furthermore, the strong modulation of the carrier density through the EDL structure enables exploration into new parameter regimes that cannot be accessed by standard solid-state metal gate structures. Superconductivity,[5, 6] as well as metal-insulator transitions[7] have been induced at electrolyte-insulator interfaces. Also, optical and electric properties in graphene have been explored over a wide range of the carrier densities.[8-12]

A drawback of the electrolyte gating is the introduction of disorder. Namely, direct deposition of an ionic liquid to the surface of interest contaminates it. While insertion of a thin boron nitride layer in between suppresses disorder effects,[13] the application of this technique is limited to placing a boron nitride flake on small devices. For a wide range of applications, quantitative evaluation of the disorder effect is necessary.

In this work, we investigate the effects of an ionic liquid on the carrier transport in graphene. The dependence of the mobility on the carrier density and temperature reveals the origin and degree of disorder from the ionic liquid. By comparing low-temperature mobility in ion- and metal-gated graphene Hall bar devices, we show that the ionic liquid introduces charged impurities, which sets the upper limit of the mobility to $3 \times 10^3$ cm$^2$/(Vs). We also show that for the higher temperature regime (> 50 K), charge carriers in graphene couple to phonons in the ionic liquid, which further reduce the mobility.

We used graphene grown by thermal decomposition of a 6H-SiC(0001) substrate.[14] Two different types of samples were fabricated; one using an ionic liquid gate and the other having a metal gate for reference.[15] Both utilized a standard Hall bar geometry with the length-to-width ratio of 3 to 1. For the ion-gated samples, the graphene Hall bar and a coplanar Cr/Au gate electrode were coated by 1-ethyl-3-methylimidazolium bis-(trifluoro-methylsulfonyl)-imide (EMIM-TFSI)[16] [Fig. 1(a)]. By applying a bias $V_{IL}$ between the gate electrode and the graphene, EDLs are formed and charge carriers are induced in graphene [Fig. 1(b)]. For the metal-gated sample, graphene was covered with 100-nm-thick hydrogen silsesquioxane and 60-nm-thick SiO$_2$ insulating layers and then the Cr/Au gate was deposited on top. The carrier density can be tuned by the top-gate bias $V_{metal}$.

We measured the longitudinal resistivity ($\rho_{xx}$) and Hall resistance ($R_{xy}$) in a top-loading cryostat with the base temperature of 4 K. The magnetic field $B$ was applied perpendicular to the graphene sample.

Since the glass transition temperature of the ionic liquid is approximately 200 K,[17] we changed $V_{IL}$ at room temperature each time before carrying out the low-temperature transport measurements.

Figure 1 (c) shows a comparison of $\rho_{xx}$ vs $B$, measured at 4 K, before and after the ionic liquid was deposited onto the sample. Prior to the deposition, the carrier density of graphene is determined by electron doping through interactions with the SiC substrate.[16] $\rho_{xx}$ shows Shubnikov-de Hass oscillations and $R_{xy}$ exhibits signs of plateaus. The carrier density and mobility were estimated to be $1.8 \times 10^{12}$ cm$^{-2}$ and 4200 cm2/(Vs) from $n = B/eR_{xy}$ and $\mu = 1/ne\rho_{xx}$ ($B = 0$ T) respectively, where $e$ is the electron charge. Simply depositing the ionic liquid to the sample, while keeping $V_{IL} = 0$ V, causes the Shubnikov-de Hass oscillations and $R_{xy}$ plateaus to essentially disappear. This indicates that the disorder is induced by the ionic liquid deposition. With this, the mobility becomes 2500 cm$^2$/(Vs), degraded by roughly half compared to the bare graphene while the density is increased to $2.75 \times 10^{12}$ cm$^{-2}$. It is worth noting that the degradation of the mobility and the increase in the density occur in all five samples we tested; the changes in the mobility and density from the ionic liquid deposition range from 34 – 62% and 120 – 190%, respectively.

To quantitatively investigate the disorder effects from the ionic liquid further, we compare transport properties for the ion- and metal-gated samples through a wide range of the carrier densities.[18] Figure 2(a) shows the Hall coefficient $R_H = R_{xy}/B$ for the ion-gated sample as a function of $V_{IL}$. The sign of $R_H$ changes at the charge neutrality point (CNP), while it varies as $1/n$ away from the CNP. Between the $R_H$ maximum and minimum, where electron and hole puddles are created, the density cannot be determined due to the potential fluctuations.[19] The minimum electron (hole) density estimated from the $R_H$ maximum (minimum) is $4.7 \times 10^{11}$ cm$^{-2}$ ($6.2 \times 10^{11}$ cm$^{-2}$). For the metal-gated sample on the other hand, $R_H$ follows $1/n$ down to $|n|\sim 1.5 \times 10^{10}$ cm$^{-2}$ [Fig. 2(b)], which is more than one order of magnitude smaller than the values for the ion-gated sample. This indicates that large potential fluctuations in the ion-gated sample primarily caused by the ionic liquid deposition.

Insight into the origin of the disorder can be gained by the $n$ dependence of the conductivity $\sigma = 1/\rho_{xx}$ [Fig. 2(c)]. Conductivity limited by defect scattering $\sigma_{def}$ is independent of $n$, whereas conductivity limited by charged impurity scattering $\sigma_{imp}$ depends linearly on $n$[20, 21] with the slope inversely proportional to the charged impurity density $n_{imp}$, $\sigma_{imp} = C \frac{e^2}{h} \frac{n}{n_{imp}}$. $C$ is a constant determined by the dielectric constant and a distance $d$ between graphene and charged impurities.[23] For the ion-gated sample, $\sigma(n)$ is linear, indicating that the charged impurity scattering is dominant. We suggest that potential fluctuations created by ions, which are roughly 1 nm away from graphene, are the principal origin of the scattering. To estimate $n_{\text{imp}}$, we use $C = 69$[22], assuming $d = 1$ nm. As a result, the linear fitting gives $n_{\text{imp}} = 7.0 \times 10^{12}$ cm$^{-2}$. For the metal-gated sample, $\sigma(n)$ shows sublinear behavior with the slope steeper around the CNP, indicating that the sample has smaller $n_{\text{imp}}$ and defect scattering plays a more dominant role. The fitting by $\sigma(n) = (\sigma_{\text{imp}}^{-1} + \sigma_{\text{def}}^{-1})^{-1}$ with $C = 69$[24] gives $n_{\text{imp}} = 1.3 \times 10^{12}$ cm$^{-2}$, which mostly comes from graphene/SiC interface states.[24] Comparison of the results for the ion- and metal-gated samples demonstrates that ions in the ionic liquid effectively serve as charged impurities with the density of $\sim 6 \times 10^{12}$ cm$^{-2}$ located at $d = 1$ nm.

As shown in Fig. 2 (d), the electron mobility in the ion-gated sample is almost constant at $\sim 2500$ cm$^2$/(Vs). This is primarily limited by the charged impurities in the ionic liquid. Specifically, $n_{\text{imp}} = 6 \times 10^{12}$ cm$^{-2}$, sets the upper limit of the mobility, $\mu_{\text{imp}} = \frac{\sigma_{\text{imp}}}{ne} = \frac{Ce}{hn_{\text{imp}}}$ to $3 \times 10^3$ cm$^2$/(Vs). Note that the hole mobility is lower than the electron mobility for the ion-gated sample and even decreases with increasing hole density. We confirmed that the data are reproducible, indicating that chemical reactions did not occur during the measurements and thus are not the origin of the electron-hole asymmetry. Although the origin is unknown, we speculate that it is related to the asymmetry of the impurities' charge,[20,25,26] which is negatively polarized around the CNP.

Finally, we investigate the effects of phonons. Figure 3 shows a comparison of the mobility as a function of temperature for the samples with and without ionic liquid.[27] For both cases, the mobility is almost constant below $T = 50$ K, above which it steadily decreases.[28] Without ionic liquid, the data can be explained by[29]

$$\mu = (\mu_0^{-1} + \mu_{\text{LA}}^{-1} + \mu_{\text{sub}}^{-1})^{-1}. \qquad (1)$$

$\mu_0 = 3.7 \times 10^3$ cm$^2$/(Vs) denotes the mobility limited due to scattering by charged impurities and defects. The contribution of LA phonons $\mu_{\text{LA}} \propto T^{-1}$ is negligibly small throughout our experimental range.[30] The decrease in the mobility at high temperature is due to phonons in the substrate, given by $\mu_{\text{sub}} = A_{\text{sub}}[\exp(E_{\text{sub}}/T) - 1]$, where $A_{\text{sub}} = 2.0 \times 10^3$ cm$^2$/(Vs) and $E_{\text{sub}} = 360$ K are obtained by the fitting. It has been shown that the contribution of phonons in the substrate on the mobility $\mu_{\text{sub}}$ does not depend on $n$.[29,30] The deposition of the ionic liquid also does not affect either $\mu_{\text{LA}}$ or $\mu_{\text{sub}}$ however, Eq. (1), with $\mu_0 = 1.6 \times 10^3$ cm$^{-2}$/(Vs) overestimates the mobility at high temperature after the deposition (blue dashed line). Phonons in the ionic liquid account for the difference and the term $\mu_{\text{IL}} = A_{\text{IL}}[\exp(E_{\text{IL}}/T) - 1]$ should be incorporated into Eq. (1). The fitting (red dashed line) gives $A_{\text{IL}} = 3.9 \times 10^3$ cm$^2$/(Vs) and $E_{\text{IL}} = 200$ K. As a result, the deposition of the ionic liquid sets the upper limit of the room temperature mobility to $\left(\mu_{\text{imp}}^{-1} + \mu_{\text{IL}}^{-1}\right)^{-1} \sim 2 \times 10^3$ cm$^2$/(Vs).

In conclusion, we evaluated the degree of disorder introduced by the electrolyte gating structure using graphene, in which the carrier transport is sensitive to disorder scattering. We found that the deposition of EMIM-TFSI introduces charged impurities, by which the low-temperature mobility in graphene is limited to $3 \times 10^3$ cm$^2$/(Vs). Above $T \sim 50$ K, phonons in the ionic liquid as well as the charged impurities further degrade the mobility. Since the degree of disorder does not depend on the base material, our results provide useful information on estimating disorder effects in any type of electrolyte gating experiments.

Acknowledgements
We thank M. Ueki for experimental support and H. Hibino for fruitful discussions.

**Fig. 1.** (a) Schematic top view of the ion-gated sample. The length and width of the Hall bar is 450 and 150 μm, respectively. (b) Schematic cross-section of the graphene EDLs. (c) $\rho_{xx}$ as a function of $B$ before (black) and after (red) the ionic liquid was applied to the sample. The gate bias remained fixed to zero after the ionic liquid application ($V_{IL} = 0$ V). The inset shows $R_{xy}$ as a function of $B$ for the same parameters.

**Fig. 2.** (a), (b) Hall coefficient $R_H = R_{xy}/B$ for the ion- and metal-gated samples as a function of $V_{IL}$ and $V_{metal}$, respectively. Note that the scales of the axes in (a) and (b) are different. The insets show the calculated carrier densities. (c) and (d) Conductivity and mobility, respectively, as a function of the carrier density for the ion-gated sample (red dots) and metal-gated sample (blue dots). The solid black lines in (c) represent results of the fitting.

**Fig. 3.** Mobility as a function of temperature before (black solid line) and after (red solid line) the deposition of the ionic liquid for $V_{IL} = 1.0$ V ($n = 6.1 \times 10^{12}$ cm$^{-2}$). The black and blue dashed lines represent the fitted data from Eq. (1) before and after the application, respectively. Incorporation of the factor $\mu_{IL}$ gives better fitting (red dashed line) for the data after the application.

Figure 1

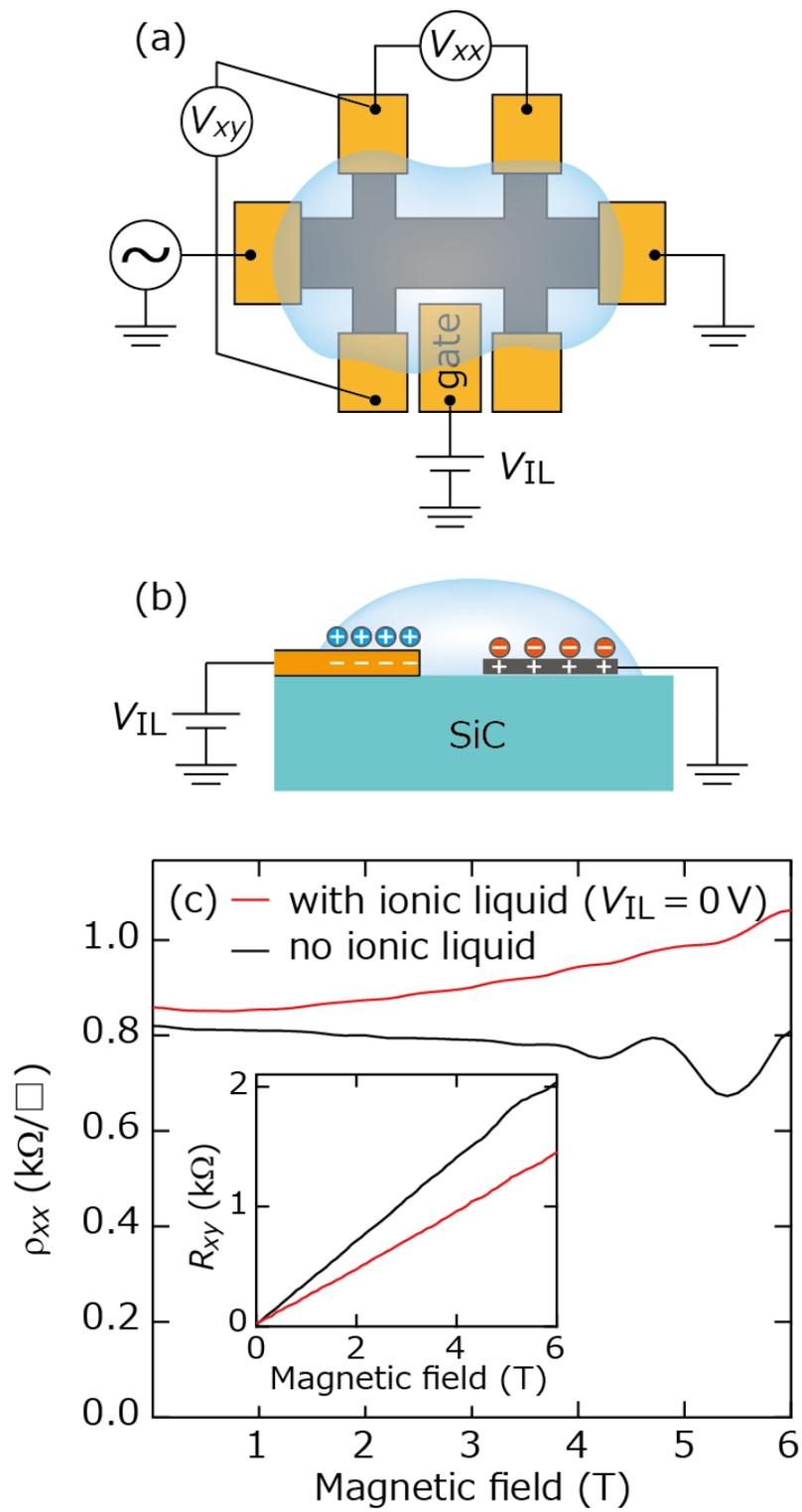

Figure 2

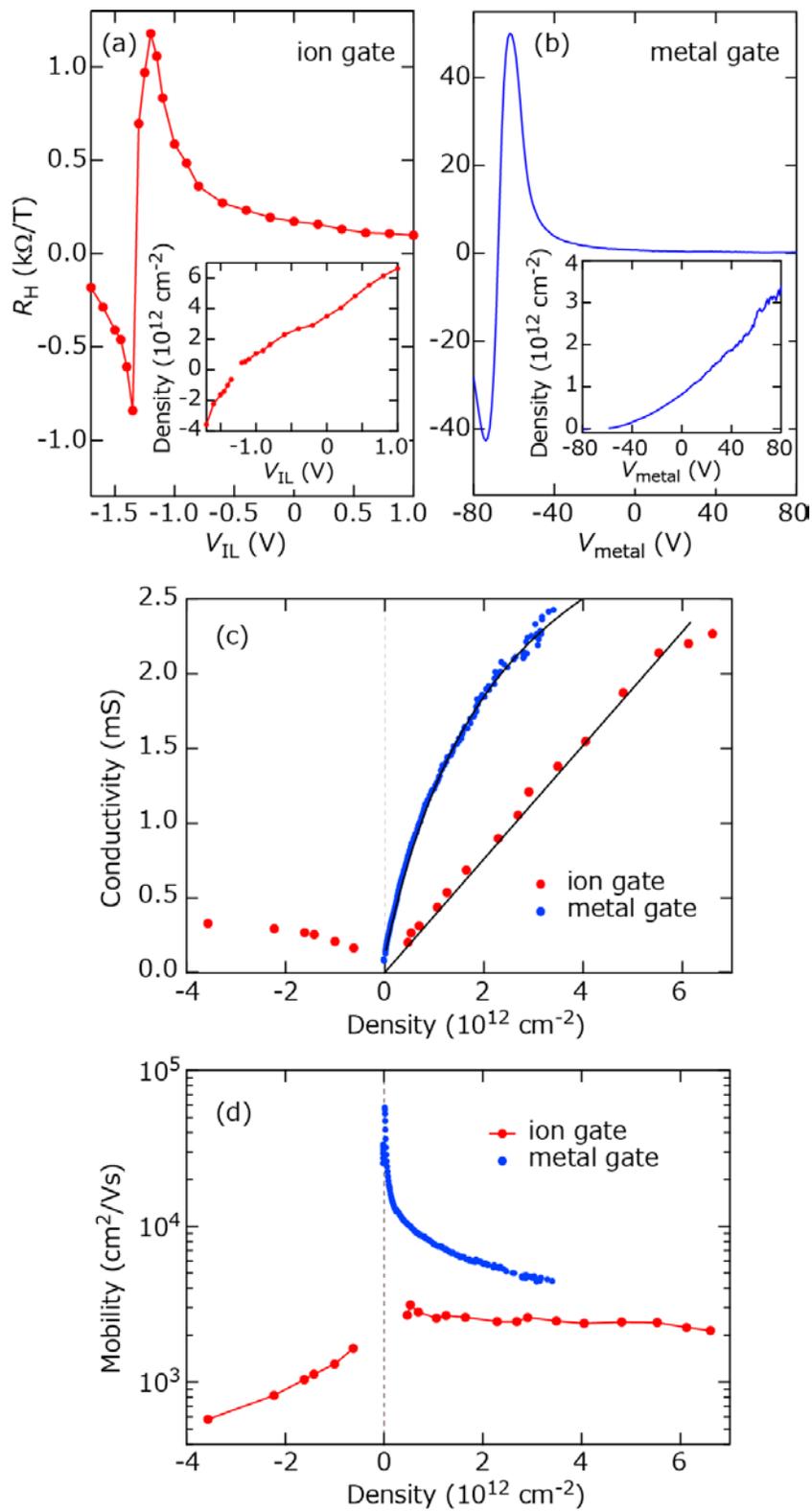

Figure 3

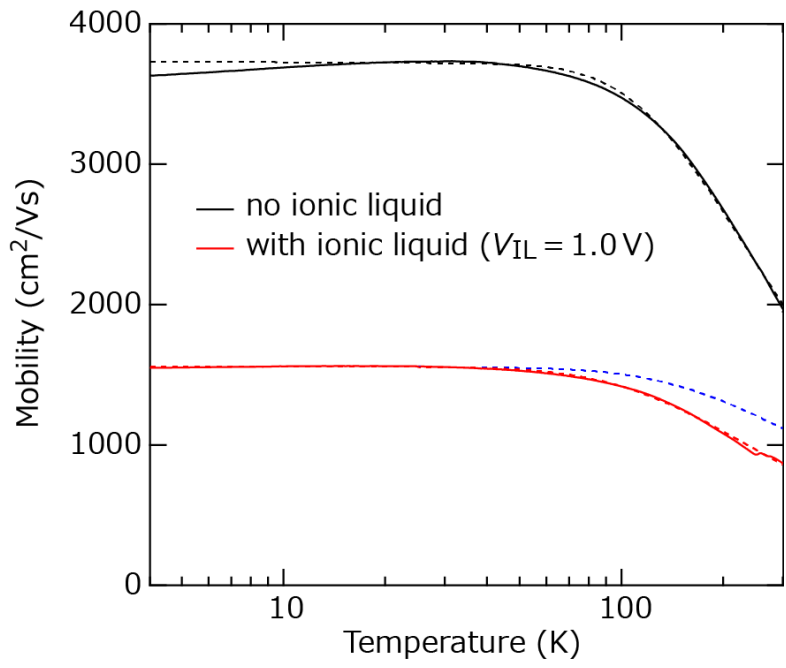

Supplementary information

In the ion-gated sample, the hole conductivity is lower than the electron conductivity at the same density magnitudes with the difference increasing with density. Here we discuss possible origins of the electron-hole asymmetry.

The sublinear behavior of the hole conductivity as a function of *n* suggests that the defect scattering plays a more dominant role for holes. However, fitting the data by $\sigma(n) = \left(\sigma_{\text{imp}}^{-1} + \sigma_{\text{def}}^{-1}\right)^{-1}$ [dashed line in Fig. S1(a)] gives $\sigma_{\text{def}} = 0.45$ mS, which is approximately one order of magnitude smaller than that used for the electron conductivity in the metal-gated sample ($\sigma_{\text{def}} = 3.8$ mS). Introduction of such strong defects just by the deposition of the ionic liquid is unlikely. Also, as mentioned in the main manuscript, chemical reactions are not the cause of the asymmetry.

One possible explanation is the asymmetry of the impurities' charge. Since the CNP is located at a negative gate bias [Fig. S1(b)], ions near the interface to graphene are negatively polarized around the CNP. The polarized charged impurities have two potential effects on the electron-hole asymmetry and also the sublinear behavior of the hole conductivity. The one is the change in the average distance *d* between charge carriers in graphene and charged impurities [1]: *d* for holes is expected to be smaller than that for electrons when the charged impurity is negatively polarized, resulting in lower conductivity for holes. This effect can also lead to the sublinear behavior if *d* decreases with decreasing gate bias. However, quantitatively, the change in the conductivity with the change in *d* by a few angstroms is expected to be only about 10% or less, not strong enough to explain the observed large electron-hole asymmetry. The other is the asymmetric scattering depending on the polarity of the impurity's charge: charge carriers in graphene are scattered more strongly when they are attracted to a charged impurity than when they are repelled from it [2]. In our case, since the polarization of the charged impurity negatively increases with decreasing the gate bias, this effect can lead to the sublinear hole conductivity. To test this speculation quantitatively, further experiments using graphene devices with different CNP gate bias values are necessary.